\documentclass{pasj00}

\begin{document}
\SetRunningHead{Saitoh \& Makino}{FAST}
\Received{2009/8/10}
\Accepted{2010/1/13}

\title{FAST: A Fully Asynchronous Split Time-Integrator for Self-Gravitating Fluid}

\author{Takayuki \textsc{R.Saitoh}$^{1}$ and Junichiro \textsc{Makino}$^{1,2,3}$
}

\affil{$^1$ Division of Theoretical Astronomy, National Astronomical
Observatory of Japan, 2--21--1 Osawa, Mitaka-shi, Tokyo 181--8588.}

\affil{$^2$ Center for Computational Astrophysics, National Astronomical
Observatory of Japan, 2--21--1 Osawa, Mitaka-shi, Tokyo 181--8588}

\affil{$^3$ Department of Astronomical Science, School of Physical Sciences,
The Graduate University for Advanced Studies (SOKENDAI), 2--21--1 Osawa,
Mitaka-shi, Tokyo 181--8588, Japan.}

\email{saitoh.takayuki@nao.ac.jp,saitoh.takayuki@cfca.jp}

\KeyWords{galaxies:starburst --- galaxies:ISM --- ISM:structure --- method:numerical}

\maketitle

\begin{abstract}
We describe a new algorithm for the integration of self-gravitating
fluid systems using SPH method. We split the Hamiltonian of a
self-gravitating fluid system to the gravitational potential and others
(kinetic and internal energies) and use different time-steps for their
integrations. The time integration is done in the way similar to that used
in the mixed variable or multiple stepsize symplectic schemes.  We performed
three test calculations. One was the spherical collapse and the other was an
explosion. We also performed a realistic test, in which the initial model
was taken from a simulation of merging galaxies.  In all test calculations,
we found that the number of time-steps for gravitational interaction were
reduced by nearly an order of magnitude when we adopted our integration
method. In the case of the realistic test, in which the dark matter
potential dominates the total system, the total calculation time was
significantly reduced.  Simulation results were almost the same with those
of simulations with the ordinary individual time-step method. Our new method
achieves good performance without sacrificing the accuracy of the time
integration.
\end{abstract}

\section{Introduction}

The number of particles used in simulations of galaxy formation with
$N$-body/Smoothed Particle Hydrodynamics (SPH) method has not increased much
since the early days of \citet{KatzGunn1991} and \citet{NavarroBenz1991},
though the number of particles used in pure $N$-body cosmological
simulations has increased drastically.  For $N$-body simulations, the
largest run in 1991 used $\sim 2 \times 10^6$ particles
\citep{SutoSuginohara1991} and the largest run recently performed used $\sim
7 \times 10^{10}$ particles \citep{Kim+2009}.  The number of particles has
grown by nearly four orders of magnitudes in two decades.  On the other
hand, for $N$-body/SPH simulations of galaxy formation, the first
simulations used $\sim 4000$ SPH particles for a single halo
\citep{KatzGunn1991} and the largest simulation which is performed recently
used $\sim 3.2 \times 10^5$ SPH particles for a single halo
\citep{Governato+2009}. {\footnote {Note that a part of SPH particles were
converted into star particles, thus the number of SPH particles was reduced
during the galaxy evolution.}} The scale up factor is only $80$ in two
decades.  This is because time-steps become quite short in dense and compact
self-gravitating gas clouds of star-forming regions.

This problem is severer in simulations with higher resolution, since these
simulations resolve denser gas.  In general, supernova (SN) explosion in
dense regions leads the shortest time-step.  Here we roughly estimate the
decrease of the time-steps due to SNe.  We consider a compact region of the
interstellar medium (ISM) with the temperature of $T_{\rm ISM}$, where the
sound speed is $c_{\rm ISM}$, as a potential site of the star formation and
that the region is rapidly heated to $T_{\rm SN}$, where the sound speed is
$c_{\rm SN}$, by SN with the energy of $E_{\rm SN}$.  The contraction factor
between the time-step of the ISM after the SN, $dt_{\rm SN}$, and before the
SN, $dt_{\rm ISM} $, in the compact region is
\begin{eqnarray}
dt_{\rm SN}/dt_{\rm ISM} &=& c_{\rm ISM}/c_{\rm SN}, \nonumber \\
&\propto& (T_{\rm ISM}/T_{\rm SN})^{1/2}, \nonumber \\
&\propto& {E_{\rm SN}}^{-1/2}~m^{1/2}~{T_{\rm ISM}}^{1/2}, \label{eq:dtratio}
\end{eqnarray}
where $E_{\rm SN}$ is the energy of the single SN and $m$ is the mass of the
heated region or the mass resolution in Lagrange schemes such as SPH,
respectively, and we use $T_{\rm SN} \propto E_{\rm SN}/m$.  From this
equation, we can easily find that the contraction factor becomes smaller
when (i) mass resolution becomes higher and (ii) the temperature of the ISM
becomes lower (see also section \ref{sec:Estimation} for more detailed
discussion). Thus high-resolution simulations which model the ISM with low
temperature ($<10^4~{\rm K}$) require much shorter time-steps than
conventional simulations of galaxy formation with a cooling cut off at $\sim
10^4~{\rm K}$.

The individual time-step method \citep{Aarseth1963, McMillan1986,
Makino1991IndividualTimeStep} reduces the total calculation cost
significantly in simulations which cover a wide range of timescales, by
assigning different time-steps to different particles and integrating only a
small fraction of particles with small time-steps.  Here, we extend this
idea for the time integration of self-gravitating fluid particles in order to
achieve a further reduction of the total calculation cost.  Our new method
allows an individual fluid particle to have different time-steps for
gravitational and hydrodynamical interactions and integrates these
interactions asynchronously.  As stated earlier, the smallest time-steps are
associated with particles heated by SNe feedback. These particles have the
thermal and kinetic energy many orders of magnitudes larger than the
gravitational potential energy.  Therefore, if we assign different
time-steps to gravitational and hydrodynamical forces, we should be able to
use much longer time-step for gravity, thereby accelerating simulations by a
large factor.  We named this time-integration scheme for self-gravitating
fluid as FAST ({\bf F}ully {\bf A}synchronous {\bf S}plit {\bf
T}ime-integrator).

There are two main advantages of the FAST method over the traditional
individual time-step method for self-gravitating fluid simulations.
First, FAST reduces unnecessary gravitational force
evaluations in small time-steps induced by SNe.  Since the number of
dark matter and stellar particles is usually larger than that of SPH
particles in typical simulations of galaxy formation, the calculation cost of
gravity is larger than that of hydrodynamics. This reduction of unnecessary
evaluation of gravity is quite efficient for the acceleration of
simulations.  Simulations with hardware accelerators, such as GRAPEs
\citep{Sugimoto+1990, Ito+1991, Okumura+1993, Makino+1997, Kawai+2000,
Makino+2003}, receive further benefit from FAST, since hardware accelerators
are inefficient in calculations with small number of particles, because of
small bandwidth and large latency of the bus between the host computer and
the accelerator.
The second advantage appears when we combine the individual time-steps and
the tree method \citep{BarnesHut1986}. 
Since the cost of the tree construction is independent of the number of
active particles, it dominates the total calculation cost when the number of
particles with small time-steps is small.  Consequently, the total performance
of simulation is not much improved by the use of individual time-steps.
Table 1 of \citet{Wadsley+2004} showed such a bad case.  We can see that a
half of the cost of smallest steps is that of the ``Tree building'' part.
By adopting FAST, the number of tree construction is reduced and hence
simulations with the individual time-step and tree methods are accelerated
significantly.

There are several other ways to reduce the cost of simulations with the
individual time-steps and tree methods.  \citet{McMillanAarseth1993} applied
the local update to the tree structure around particles which were updated,
instead of reconstructing the whole tree structure at each step. This
technique was used in {\tt GADGET-1/2} \citep{Springel+2001, Springel2005}.
In {\tt VINE} \citep{Wetzstein+2009, Nelson+2009}, the construction
frequency of tree structure was reduced by skipping several continuous
time-steps and reusing old tree structure for force calculation.  They
updated the tree structure at every $\sim 10$ steps for the problem they
showed in their paper.  FAST method can be combined with these schemes to
further reduce the cost of tree construction, if necessary.

Our approach is similar to the multiple time-step method used in molecular
dynamics, in which the long-range Coulomb force is updated less frequently
than short-range van del Waals force \citep{Streett+1978}. The main
difference is that we combine the force splitting with individual
time-steps.

The structure of this paper is as follows.  In section \ref{sec:Estimation},
we estimate and compare the time-steps of particles in the hot region of
star-forming galaxies.  In section \ref{sec:Idea}, we describe our new
integration method for self-gravitating fluid, FAST.  We briefly explain its
implementation in \S \ref{sec:Implimentation}.  We present the results of
test calculations and timing results in section \ref{sec:Tests}.  A
discussion on the maximum acceleration factor by FAST appears in section
\ref{sec:Discussion}.  In section \ref{sec:Summary}, we provide summary.

\section{Estimate of Time-steps in Heated Regions of Star Forming Galaxies} \label{sec:Estimation}

In this section, we estimate typical time-steps of an SPH particle in star
forming regions of actively star forming galaxies in $N$-body/SPH
simulations of galaxy formation. This estimation allows us to estimate the
maximum gain in the performance due to the use of the FAST scheme. We
compare the typical Courant time-step of an SPH particle heated by SNe with
the typical gravitational time-step of the particle.

Here we estimate the typical Courant time-step of an SPH particle heated by
SNe.  For simplicity, we adopt following four assumptions.  First, we adopt
a single stellar population (SSP) approximation for a star particle with
Salpeter initial mass function (IMF) \citep{Salpeter1955} and the range of
this IMF is set to be $0.1~\Mo$ to $100~\Mo$.  For this IMF, the specific SN
rate is $\epsilon_{\rm SN} \simeq 0.0074~{\rm SN/\Mo}$, where we assume
$8~\Mo$ or heavier stars become SNe at the final phase of their evolutions.
Second, we assume that each SN injects the thermal energy of $E_{\rm SN} =
10^{51}~{\rm ergs}$ to the surrounding ISM (the nearest $N_{\rm NB}$
particles).  Third, we assume that the whole energy from SNe in a star
particle discharges in a single event (this is one of SN feedback
implementations proposed by \cite{Okamoto+2008}).  Finally, we assume that
the masses of the stellar and gas particles are the same.

The mean additional internal energy for $N_{\rm NB}$ SPH particles due to
SNe of a single stellar particle is given by 
\begin{eqnarray}
U_{\rm SN} &=& \frac{\epsilon_{\rm SN} m_{*} E_{\rm SN} }{N_{\rm NB} m_{\rm SPH}}, \nonumber \\
&\simeq& 0.0074 \times 10^{51} \times \frac{m_{*}}{N_{\rm NB} m_{\rm SPH}}~[{\rm ergs~\Mo^{-1}}], \nonumber \\
&\simeq& \frac{3.7 \times 10^{15}}{N_{\rm NB}} [{\rm ergs~g^{-1}}], 
\end{eqnarray}
where $m_{*}$ and $m_{\rm SPH}$ are the masses of stellar and gas particles,
respectively, and we use the relation $m_{*} = m_{\rm SPH}$.  The sound
speed, $c_{\rm SN}$, of the heated gas region is 
\begin{eqnarray}
c_{\rm SN} &\simeq& \sqrt{\gamma (\gamma -1) U_{\rm SN}}, \nonumber \\
&\simeq& \frac{6.4 \times 10^2}{{N_{\rm NB}}^{1/2}}~[{\rm km~s^{-1}}], \label{eq:soundspeed}
\end{eqnarray}
where we assume an ideal gas with the adiabatic index of $\gamma = 5/3$.
The original internal energy of the ISM before SNe, $U_{\rm ISM}$, is quite
small, therefore we neglected $U_{\rm ISM}$ in the estimation of $c_{\rm
SN}$. 
The corresponding temperature of the heated region is $T_{\rm SN} \sim 3.2
\times 10^{6}~{(N_{\rm NB}/32)}^{-1/2}~{\rm [K]}$.  
Note that this temperature implies very short cooling timescale of $\sim
10^3~{\rm yr}$.  In real star-forming region, initially the SN ejecta have
much high temperature, and the cooling time is much longer.  In order to
model SN feedback in a physically correct way, therefore, some tricks which
prevent the quick radiative cooling \citep{Gerritsen1997PhD,
ThackerCouchman2000, Stinson+2006} is necessary. We here assume some of
these tricks are used.
The size of an SPH particle, $\lambda$, is 
\begin{equation}
\lambda = \Bigl ( \frac{3}{4 \pi} \frac{m_{\rm SPH}}{\rho} \Bigl )^{1/3}, \label{eq:size}
\end{equation}
where $\rho$ is the density of the SPH particle.  Combining equations
(\ref{eq:soundspeed}) and (\ref{eq:size}), we obtain the sound crossing
time, $t_{\rm SN} \equiv \lambda/c_{\rm SN}$, in the heated region as
follows:
\begin{eqnarray}
t_{\rm SN} \simeq 4 \times 10^4 \Bigl ( \frac{{m_{\rm SPH}}}{1000~\Mo} \Bigl )^{1/3}
\Bigl ( \frac{100~{\rm cm^{-3}}}{N_{\rm H}} \Bigl )^{1/3}
[{\rm yr}], \label{eq:t_SN}
\end{eqnarray}
where we adopt $N_{\rm NB} = 32$ and $N_{\rm H}$ is the hydrogen number
density of the heated region.  The typical Courant time-step in the region is
$dt_{\rm SN} \sim 0.1 \times t_{\rm SN}$.  This equation tells us that the
smallest time-step in simulations involving the low temperature ISM and SNe
becomes shorter when mass resolution becomes higher and injected region
becomes denser.

By comparing equation (\ref{eq:soundspeed}) with the typical velocity of the
ambient ISM, we can obtain the contraction factor of the time-steps caused
by a SN explosion.
Although the typical temperature of giant molecular clouds (GMCs) is low
($\sim 10~{\rm K}$) and the corresponding sound speed in GMCs is small
($\sim 0.2~{\rm km~s^{-1}}$), the linewidth of GMCs is higher than that
expected by the sound speed of the ISM and predicts that GMCs are supported
by supersonic turbulence.  Thus we use empirical relations for the estimate
of the timescale, instead of the local sound speed.  The linewidth-size
relation, which is often referred as Larson's law \citep{Larson1981,
Solomon+1987, HeyerBrunt2004}, gives us the typical velocity at the size of
cloud.  Larson's law is as follows:
\begin{equation}
\sigma_{\rm c} \simeq \Bigl ( \frac{L_{\rm c}}{1~{\rm pc}} \Bigl )^{1/2}~[{\rm km~s^{-1}}], \label{eq:larson}
\end{equation}
where $\sigma_{\rm c}$ and $L_{\rm c}$ are the linewidth and size of a
cloud, respectively, and the applicable range of this relation is $0.1~{\rm
pc} \le L_{\rm c} \le 100~{\rm pc}$. Combining the virial theorem and this relation,
we obtain cloud mass-linewidth relation \citep{Solomon+1987} that
\begin{equation}
M_{\rm c} = 2000 \Bigl ( \frac{\sigma_{\rm c}}{1~{\rm km~s^{-1}}} \Bigl )^4~\Mo,
\end{equation}
where $M_{\rm c}$ is a cloud mass.  When we substitute $N_{\rm NB} m_{\rm
SPH}$ into $M_{\rm c}$, we obtain the mass resolution-linewidth relation:
\begin{equation}
{\sigma_{\rm c}} = \Bigl ( \frac{N_{\rm NB} m_{\rm SPH}}{2000~\Mo}\Bigl )^{1/4}~{\rm [km~s^{-1}]}. 
\label{eq:resolution-linewidth}
\end{equation}
This equation leads the velocity of the smallest cloud which can be
expressed with the resolution of the simulation.  The contraction factor of
the time-step in the ISM by a SN explosion, $f_{\rm
cont}$, is 
\begin{eqnarray}
f_{\rm cont} &\equiv& \frac{\sigma_{\rm c}}{c_{\rm SN}}, \nonumber \\
&=& \Bigl ( \frac{N_{\rm NB} m_{\rm SPH}}{2000~\Mo}\Bigl )^{1/4} \Bigl ( \frac{6.4 \times 10^2}{{N_{\rm NB}}^{1/2}} \Bigl )^{-1}, \nonumber \\
&\sim& 1.8 \times 10^{-2} \Bigl ( \frac{m_{\rm SPH}}{1000~\Mo} \Bigl )^{1/4},
\label{eq:r_cont}
\end{eqnarray}
where we adopted $N_{\rm NB} = 32$. This equation clearly shows that the
Courant condition becomes quite tight in ISM heated by a SN explosion.
When we use $6~{\rm km~s^{-1}}$, which is the sound speed of the ISM at
$10^4~{\rm K}$, as the typical velocity of the ISM, the contraction factor is
$f_{\rm cont} \sim 5.3 \times 10^{-2}$. We again adopted $N_{\rm NB} = 32$.
The contraction factor for simulations with the multiphase ISM and turbulence
motions is smaller than that in conventional simulations of galaxy formation
with a cooling cut off at $10^4~{\rm K}$.

In conventional simulations of galaxy formation, where the typical mass
resolution is $10^6~\Mo$ and the highest density of the ISM is $0.1~{\rm
cm^{-3}}$, the typical time-step for the heated region is $dt_{\rm SN} \sim
4 \times 10^5~{\rm yr}$. The typical gravitational time-step, one-tenth of
the local free-fall time at $0.1~{\rm cm^{-3}}$, is $5\times10^6~{\rm yr}$.
The difference between the Courant and gravity time-steps is $\sim 10$.
On the other hand, in state-of-the-art simulations involving the multiphase
ISM, where $m_{\rm SPH} = 10^3~\Mo$ and $N_{\rm H} = 100~{\rm cm^{-3}}$, the
typical time-step is $dt_{\rm SN} \sim 4 \times 10^3~{\rm yr}$, whereas the
the typical gravitational time-step at $100~{\rm cm^{-3}}$ is
$1.6\times10^5~{\rm yr}$.  The difference between two time-steps is $\sim
40$ and this difference is larger than that in conventional simulations.
These simple estimates tell us that FAST can reduce gravity steps by a
factor of $10-40$.  FAST is more efficient in simulations with high resolution.

Thanks to the rapid increase of the computational power and the advance of
numerical techniques, the mass resolution in current high resolution
simulations of the galactic scale ISM has been quite high ($\sim 1000~\Mo$).
It will be soon reach the point than the mass resolution at where the number
of SN events in an SSP particle is less than unity.
{\footnote{When an SSP particle mass is lower than a critical mass $m_{\rm
c}$, which is obtained by $\epsilon_{\rm SN} m_{\rm c} = 1$, the number of
SNe events in an SSP particle is lower than unity. If we use $\epsilon_{\rm
SN} = 0.0074$, $m_{\rm c} \sim 135~\Mo$.
}}
In such simulations, SN events are necessary to be treated as not an
association of SNe in every SSP particle but discrete events in a fraction
of SSP particles so that the global SN event rate is consistent with the
adopted IMF.  Otherwise, we would introduce ``fractional'' SNe, which
clearly would give wrong results for SNe feedback.  By this modification in
the treatment of SNe, the sound speed of the ISM in heated regions becomes
much higher than that in equation (\ref{eq:soundspeed}) and the crossing
time in these regions becomes much shorter than that in equation
(\ref{eq:t_SN}). We show here a simple estimate of time-steps in the case
where SN explosions are discrete events in SSP particles.  When we consider
a SN explosion takes place in discrete manner, the received energy of the
surrounding ISM of a compact region is modified as follows:
\begin{eqnarray}
U_{\rm SN,d} &=& \frac{E_{\rm SN} }{N_{\rm NB} m_{\rm SPH}}, \nonumber \\
&\simeq& \frac{10^{51}}{N_{\rm NB} m_{\rm SPH}}~[{\rm ergs~\Mo^{-1}}], \nonumber \\
&\simeq& \frac{5.0 \times 10^{17}}{N_{\rm NB}} 
    \Bigl ( \frac{1~\Mo}{m_{\rm SPH}} \Bigl )~[{\rm ergs~g^{-1}}], 
\end{eqnarray}
where we again neglected the original internal energy because the value is
sufficiently low compared with this value.  The sound speed of the hot
region is 
\begin{eqnarray}
c_{\rm SN,d} &=& \sqrt{\gamma (\gamma -1) U_{\rm SN,d}}, \nonumber \\
&\simeq& \frac{7.5 \times 10^{3}}{{N_{\rm NB}}^{1/2}} 
\Bigl ( \frac{1~\Mo}{m_{\rm SPH}} \Bigl )^{1/2}~[{\rm km~s^{-1}}]. \label{eq:soundspeed,d}
\end{eqnarray}
The contraction factor is 
\begin{eqnarray}
f_{\rm cont} &=& \Bigl ( \frac{N_{\rm NB} m_{\rm SPH}}{2000~\Mo}\Bigl )^{1/4} 
\Bigl \{ \frac{7.5 \times 10^{3}}{{N_{\rm NB}}^{1/2}}
\Bigl ( \frac{1~\Mo}{m_{\rm SPH}} \Bigl )^{1/2} \Bigl \}^{-1}, \nonumber \\
&\simeq& 2.7 \times 10^{-4} \Bigl ( \frac{m_{\rm SPH}}{1~\Mo} \Bigl )^{3/4}.
\end{eqnarray}
Note that the mass dependency in this equation is much stronger than that in
equation (\ref{eq:r_cont}).  Combining equations (\ref{eq:soundspeed,d}) and
(\ref{eq:size}), the sound crossing time in the heated region, $t_{\rm
SN,d}$, is 
\begin{equation}
t_{\rm SN,d} \simeq  3.3 \times 10^2 
\Bigl ( \frac{m_{\rm SPH}}{1~\Mo} \Bigl )^{5/6}
\Bigl ( \frac{100~{\rm cm^{-3}}}{N_{\rm H}} \Bigl )^{1/3}~[{\rm yr}], \label{eq:t_SN,d}
\end{equation}
where we adopt $N_{\rm NB} = 32$.  We find that the mass resolution
dependence in the equation (\ref{eq:t_SN,d}) is stronger than that in the
equation (\ref{eq:t_SN}).  Thus high resolution simulations of near future
will be much harder than those of present. For efficient simulations, we
have to introduce efficient numerical techniques which can handle a very
wide range of time-steps.  We believe that our new scheme will play an
important role not only in current simulations but also in new simulations
of galaxy formation in the near future.

\section{Basic Idea} \label{sec:Idea}

The basic idea of our new scheme is as follows.  We allow gas (SPH)
particles to have different time-steps for gravitational and hydrodynamical
integrations.  Thus, we extend the idea of individual time-steps, which
allows different particles to have different time-steps, to allow single
particle to have different time-steps for different interactions.  We then
asynchronously integrate gravity and hydrodynamics with these different
time-steps.  This is the essence of our FAST method.  Since the problem we
have to solve is the time-integration of very hot gas particles formed by SNe,
we allow time-steps for gravity to be longer than those for hydrodynamics.
To allow different time-steps for gravity and hydrodynamics, we use the
technique of constructing multi-timestep symplectic integrator.  We divide
the Hamiltonian of a self-gravitating fluid system into a gravitational
potential term and others, and integrate each part with its own time-step.

The Hamiltonian of a self-gravitating fluid system of $N$ gas particles is
expressed as 
\begin{equation}
H = \sum_{i}^{N} \frac{p_i^2}{2 m_i} 
+ U({\boldsymbol q},{\boldsymbol \rho},{\boldsymbol s}) 
- \sum_{i}^{N} \frac{G m_i m_j}{q_{ij}},
\label{eq:hamiltonian_hydro}
\end{equation}
where $p_i$ and $q_i$ are conjugate variables of the canonical equation for
particle $i$, $m_i$ is the mass of particle $i$, $U$ is the internal energy
of fluid, which is a function of ${\boldsymbol q}$, density, ${\boldsymbol
\rho}$, and entropy, ${\boldsymbol s}$.  Here, ${\boldsymbol q}$,
${\boldsymbol \rho}$, and ${\boldsymbol s}$ denote $(q_1, q_2, q_3, \ldots,
q_{N})$, $(\rho_1, \rho_2, \rho_3, \ldots, \rho_{N})$, and $(s_1, s_2, s_3,
\ldots, s_{N})$, respectively.  Since we take into account arbitrary forms
of hydrodynamical interactions, we express the internal energy for fluid as
$U({\boldsymbol q},{\boldsymbol \rho},{\boldsymbol s})$.  The first, second,
and third terms in the right hand side of equation
(\ref{eq:hamiltonian_hydro}) are the kinetic, internal, and gravitational
potential energy of the system, respectively.  The actual equations for $p$
and $s$ contain the contributions of non-conservative terms like artificial
viscosity and radiative cooling/heating.  For simplicity, we here regard the
system as adiabatic (i.e., $s_i$ are treated as constants). Hence the
internal energy term becomes the function of (${\boldsymbol q},{\boldsymbol
\rho}$) and can be regarded as a potential term in the Hamiltonian.

We split the Hamiltonian into the gravitational potential term and others
(see appendix \ref{sec:SymplecticSchemes}):
\begin{eqnarray}
H_{\rm hydro} &=& \sum_{i}^{N} \frac{p_i^2}{2 m_i} 
      + U({\boldsymbol q},{\boldsymbol \rho}), \label{eq:Hydro} \\
H_{\rm grav} &=& - \sum_{i}^{N} \frac{G m_i m_j}{q_{ij}}. \label{eq:Gravity} 
\end{eqnarray}
We then obtain the following expression of a symplectic integrator with the
second-order accuracy,
\begin{equation}
f(t+\Delta t) \approx  
e^{\frac{\Delta t}{2} \{,H_{\rm grav}\}} 
e^{\Delta t \{,H_{\rm hydro}\}} e^{\frac{\Delta t}{2} \{,H_{\rm grav}\}} 
f(t), \label{eq:gravity_hydro_integral}
\end{equation}
where ``$\{ \ , \ \}$'' is a Poisson bracket and $\Delta t$ is a time-step.
The equation (\ref{eq:gravity_hydro_integral}) can schematically rewrite as
follows:
\begin{eqnarray}
v'_0 &=& v_0 + \frac{1}{2} \Delta t~a_{\rm grav}, \\
x_0 &\rightarrow& ({\rm Hydro ~update}) \rightarrow x_1, \\
v'_0 &\rightarrow& ({\rm Hydro ~update}) \rightarrow v'_1, \\
v_1 &=& v'_1 + \frac{1}{2} \Delta t~a_{\rm grav},
\end{eqnarray}
where $x$, $v$, $v'$, and $a_{\rm grav}$ indicate the position, the velocity,
the half-step advanced velocity, and the acceleration of gravitational
force, respectively. Subscripts $0$ and $1$ indicate epochs of
time-integration at $t$ and $t + \Delta t$, respectively.

There are many ways to integrate the hydrodynamical part of equation
(\ref{eq:gravity_hydro_integral}), we here choose the second-order
symplectic method (e.g., \cite{HernquistKatz1989}). We divide equation
(\ref{eq:Hydro}) again into the following two parts:
\begin{eqnarray}
H_{\rm hydro,T} &=& \sum_{i}^{N} \frac{p_i^2}{2 m_i}, \\
H_{\rm hydro,U} &=& U({\boldsymbol q},{\boldsymbol \rho}).
\end{eqnarray}
Consider the case that $\Delta t_{\rm g} = l \Delta t_{\rm h}$, where $l$ is
a natural number. We obtain a new expression of equation
(\ref{eq:gravity_hydro_integral}) as
\begin{eqnarray}
f(t+\Delta t) &\approx& 
e^{\frac{\Delta t_g}{2} \{,H_{\rm grav}\}} \nonumber \\
&& [e^{\frac{\Delta t_h}{2} \{,H_{\rm hydro,U}\}} 
    e^{\Delta t_h \{,H_{\rm hydro,T}\}} 
    e^{\frac{\Delta t_h}{2} \{,H_{\rm hydro,U}\}}]^l \nonumber \\
&& e^{\frac{\Delta t_g}{2} \{,H_{\rm grav}\}} f(t).
    \label{eq:gravity_hydro_integral_subcycle}
\end{eqnarray}
This equation tells us that we can reduce the computational cost of gravity
if $\Delta t_g > \Delta t_h$ ($l > 1$).  If we adopt $l = 1$, the integrator
is the same as the standard ``leap-frog'' method for self-gravitating
fluid.

\begin{figure*}[htb]
\begin{center}
\includegraphics[width=0.95 \textwidth]{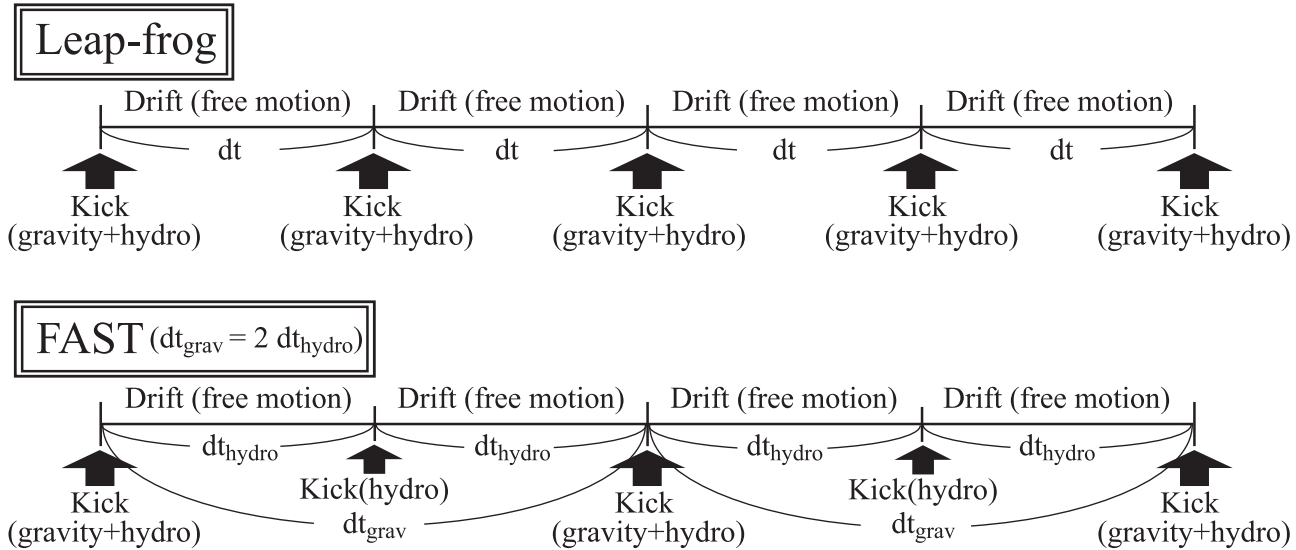}
\caption{
\label{fig:SchematicPicture}
The schematic picture of the leap-frog and FAST methods for the integration
of a self-gravitating fluid.  See also figure 1 in \citet{Fujii+2007} for
MVS and BRIDGE methods.  ``Kick'' means the momentum exchanges between
particles, while ``Drift'' denotes the free (inertial) motions under given
velocity vectors. 
}
\end{center}
\end{figure*}

In figure \ref{fig:SchematicPicture}, we show schematic pictures of the
usual leap-frog and FAST methods.  For FAST, we consider the case that
$dt_{\rm grav} = 2~dt_{\rm hydro}$.  The computational cost of gravitational
force in FAST becomes half of the leap-frog method in this case. In
practice, the time-step ratio, $l$, adaptively changes.

It should be noted that, even though we borrowed the formalism of symplectic
integrators to describe our FAST method, the FAST method itself is not
symplectic. This is because we change the time-steps for gravitational and
hydrodynamical interactions, after we split the Hamiltonian. In addition, we
use different time-steps for different particles.  However, this issue is
not as crucial as that is for pure $N$-body simulations, since the usual
hydrodynamical simulations introduce a dissipation term. The
time-integration of hydrodynamic simulations is usually time irreversible
and is breaking the symplectic nature inherently.

There have been several proposed methods which can retain either
symplecticness \citep{FarrBerschinger2007} or time symmetry
\citep{Makino+2006} when used with the individual time-step method.
However, these schemes are computationally expensive and it is not clear if
the use of these schemes is worthwhile or not.  In this paper, we
concentrate on the traditional, non-symplectic implementation of individual
time-step algorithm and its extension.

\section{Implementation} \label{sec:Implimentation}

\subsection{The Code}

The code used in this paper is a parallel tree SPH code, {\tt ASURA}, which
utilizes the special purpose hardware GRAPE (Saitoh in prep.).
Gravitational force was solved by Tree with GRAPE
\citep{Makino1991TreeWithGRAPE}.  In this paper, we used the Phantom GRAPE
library for calculations of gravity, which is a software emulator of GRAPE
pipelines (kindly provided by Kohji Yoshikawa).  We used an opening angle of
0.5 and only monopole moments for force calculations.  Hydrodynamics was
followed by the standard SPH method (e.g., \cite{Lucy1977,
GingoldMonaghan1977, Monaghan1992}).  We used the ``gather'' formulation of
SPH for the density estimation, whereas the ``gather and scatter''
formulation of SPH for the pressure gradient and the time derivation of
internal energy \citep{Monaghan1992}.  We adopted the asymmetric form energy
equation (e.g., \cite{SteinmetzMuller1993}).  We iteratively determined the
kernel radius of each SPH particle in every step in order to keep the number
of neighbor particles, $32 \pm 2$.  We used an artificial viscosity term, of
which form is the same as that proposed by \citet{Monaghan1997}, in order to
handle shocks.  The value of the viscosity parameter, $\alpha$, was set to
be unity.  {\tt ASURA} adopts the variable and individual time-step method
\citep{McMillan1986, HernquistKatz1989}.  Following
\citet{Makino1991IndividualTimeStep}, {\tt ASURA} adopts an extended version
of the individual time-step method, {\it i.e.}, the ``hierarchical''
time-step method where time-steps are quantized by the power of two of a
baseline time-step in order to improve the simulation performance with
individual time-steps.  We also implemented the time-step limiter for
hydrodynamical interactions in order to maintain the difference of
time-steps among neighbor particles small enough \citep{SaitohMakino2009}.
Here we adopted the factor of the time-step difference in neighbors, $f =
4$.  The current version of {\tt ASURA} implements two time integrators,
namely the ordinary leap-frog and FAST methods.  

\subsection{Time-steps}

The time-steps were determined as follows.  The gravity time-step of an
$i$-th particle was estimated by 
\begin{equation}
dt_{{\rm grav},i} = C_{\rm grav} 
\min  \Bigl ( \sqrt{\frac{\epsilon}{|{\boldsymbol a}_{{\rm grav},i}|}}, 
\frac{|{\boldsymbol a}_{{\rm grav},i}|}{|\dot{\boldsymbol a}_{{\rm grav},i}|} \Bigl ), \label{eq:dtgrav}
\end{equation}
where $\epsilon$ is a gravitational softening length, $C_{\rm grav}$ is a
parameter which controls the accuracy (we here adopt 0.1), and
$\dot{{\boldsymbol a}}_{{\rm grav},i}$ is the time derivation of the
acceleration, respectively.

Following \citet{Monaghan1997}, the hydrodynamical time-step of the $i$-th
SPH particle was determined by  
\begin{equation}
dt_{{\rm hydro},i} = C_{\rm hydro} \frac{2 h_i}{v_{{\rm sig},i}}, \label{eq:dthydro}
\end{equation}
where $h_i$ is the kernel size of the SPH particle (the interaction scale is
$2 h_i$), $C_{\rm hydro} = 0.25$, and $v_{{\rm sig},i}$ is the local maximum
signal-velocity of $i$-th particle defined by 
\begin{equation}
v_{{\rm sig},i} = \max_{j}(c_i + c_j - 3 w_{ij}),
\end{equation}
where $j$ indicates the indices of neighbor particles, $c_i$ and
$c_j$ is the sound speed of $i$-th and $j$-th SPH particles and $w_{ij} =
{\boldsymbol v}_{ij} \cdot {\boldsymbol x}_{ij}/|{\boldsymbol x}_{ij}|$ is a
projected relative velocity between the SPH particles.  We set $w_{ij} = 0$
if $w_{ij} > 0$.

When we use the FAST method, we asynchronously integrate gravity and
hydrodynamics by the leap-frog method with different time-steps for gravity
(Eq.  \ref{eq:dtgrav}) and hydrodynamics (Eq. \ref{eq:dthydro}).  If
$dt_{\rm grav} \neq 2^n dt_{\rm hydro}$ where $n$ is an integer number of
$\ge 0$, we change the gravitational time-step so that it satisfies the
above criterion in the following way: we reduce the time-step of gravity to
$dt_{\rm grav}'$, where $dt_{\rm grav} \ge dt_{\rm grav}' = 2^n dt_{\rm
hydro}$, and $n$ is the maximum integer number that satisfies this relation.
If $dt_{\rm grav} < dt_{\rm hydro}$, we reduce $dt_{\rm hydro}$ to the same
value as $dt_{\rm grav}$.

When we used the ordinary leap-frog method, we picked up the smaller one of
the two time-steps as the time-step of an SPH particle,
\begin{equation}
dt = {\rm min} (dt_{\rm grav},dt_{\rm hydro}), \label{eq:dts}
\end{equation}
and synchronously integrated both gravity and hydrodynamics. In general, the
acceleration and its differential terms in equation (\ref{eq:dtgrav})
should be measured relative to the total acceleration ({\it i.e.}, the sum
of the gravitational and hydrodynamical accelerations).  However, in the
hydrodynamical simulations, the Courant condition leads the smaller
time-step compared with the time-step obtained by the total acceleration.
Therefore the simple determination by equation (\ref{eq:dts}) worked
sufficiently.

\section{Numerical tests} \label{sec:Tests}

We performed three tests. The first test was the collapse of a gas cloud and
the second test was the point-like explosion of a self-gravitating gas
cloud.  The third test was a more realistic simulation.  We performed
simulations of galaxy-galaxy collisions, where galaxies consist of dark
matter, star and gas particles.  These tests incorporated both gravity and
hydrodynamics and were representative of the evolution of self-gravitating
fluid in galaxy formation simulation or other astrophysical simulations.
We, hereafter, denote the results with the ordinary individual time-step
method as ``Ind'' and the results with the individual time-step with the
FAST method as ``FAST'', in tables and figures. The first two tests were
designed as simple tests for the validity of the FAST method, while in the
third test we investigated the actual gain in the calculation speed as well
as the accuracy of the result.

The first and second tests were done on a system with a $2.4$ GHz Opteron
$280$ processor (Italy core), while the third test was done on $2.2$ GHz
quad-core Opteron processors (Barcelona core) of Cray XT4 system at Center
for Computational Astrophysics of National Astronomical Observatory of
Japan.  We used one CPU core for the first and second tests, whereas we used
$128$ CPU cores for the third test.

\subsection{Test I: Three dimensional self-gravitational collapse tests}

We performed the integration of three-dimensional spherical collapse of
adiabatic gas (e.g., \cite{Evrard1988,HernquistKatz1989}).  This test is one
of standard tests for SPH method which involves self-gravity.

We prepared a gas sphere with the total mass and the radius both unity.  The
gravitational constant was also set to be unity.  The initial profile of the
gas sphere was $\rho(r) \propto 1/r$, where $r$ is the distance from the
center of coordinates.  The adiabatic index and the specific internal energy
of the gas were set to be $\gamma = 5/3$ and $0.05$, respectively.  The gas
sphere had a negative value of the total energy, $E \sim -0.6$.  When the
evolution starts, the sphere begins to collapse.  The shock takes place in
the central region and it propagates outward.  Finally, the system reaches
the state of virial equilibrium.  In this test, we used $30976$ particles
for the sphere and we followed the evolution of the gas sphere to $T = 3$.
We adopted $0.038$ for the gravitational softening length.

\begin{figure*}[htbp]
\begin{center}
\includegraphics[width=0.95 \textwidth]{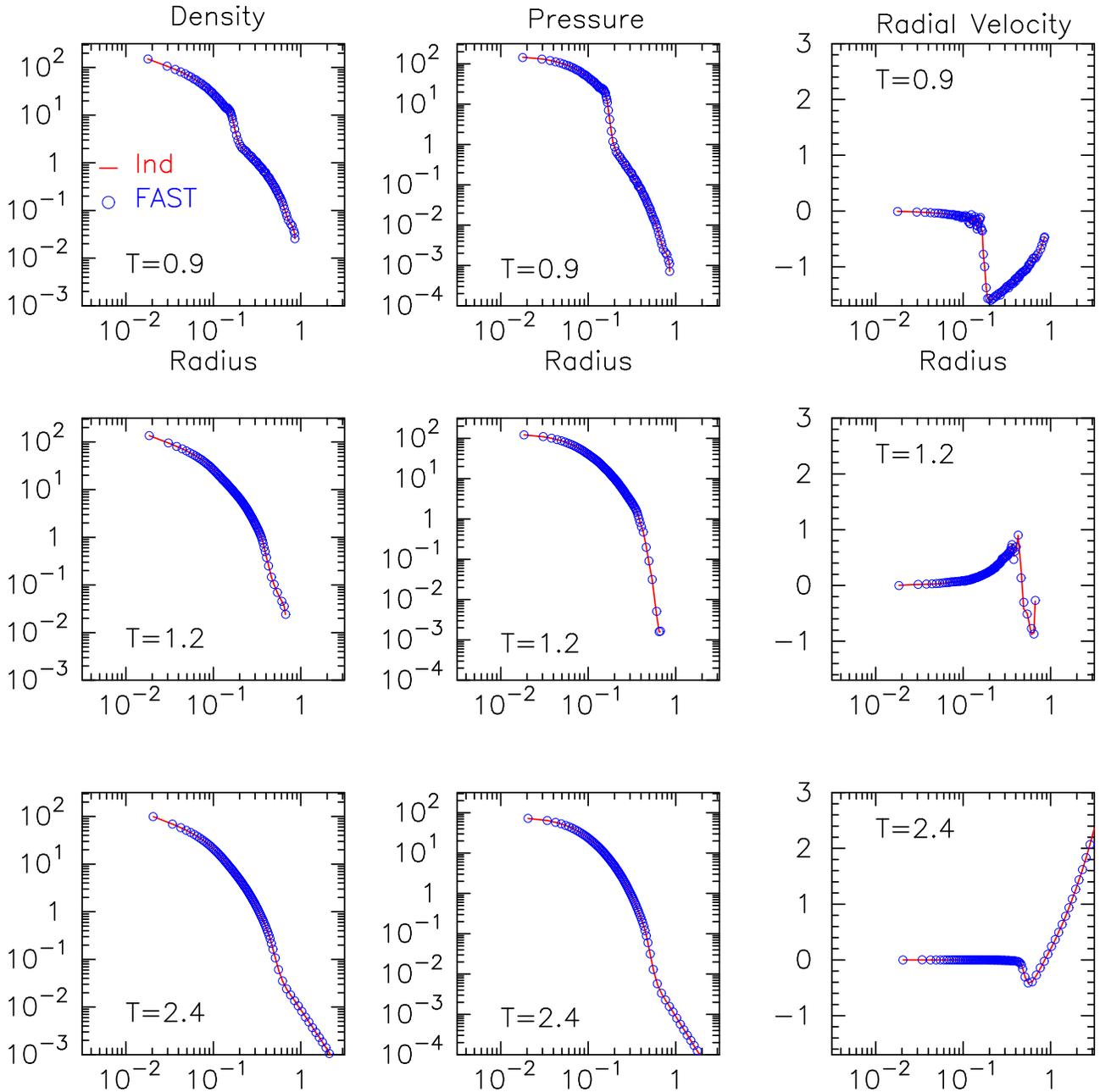}
\caption{
\label{fig:3dcollapse}
Radial profiles of density (left), pressure (mid), and radial velocity
(right) at $T = 0.9, 1.2,$ and $2.4$ (top to bottom).  Horizontal axis is
the distance from the origin of coordinates.  Curves and circles indicate
the profiles obtained with the individual and synchronous time-steps for
gravity and hydrodynamics, ``Ind'', and  the individual and asynchronous
time-steps for gravity and hydrodynamics,``FAST''.
}
\end{center}
\end{figure*}

Figure \ref{fig:3dcollapse} shows radial profiles of density, pressure, and
radial velocity in three different epochs, $T = 0.9$, $1.2$ and $2.4$, for
both the ordinary individual time-step and FAST methods.  In this figure, we
plotted mean physical quantities of every 300 particles.  It is obvious that
the results of two methods are identical.  We also confirmed that these
results agree well with the result obtained using global time-steps.
Therefore we can conclude our new method is accurate enough.

\begin{figure}[htbp]
\begin{center}
\includegraphics[width=0.45 \textwidth]{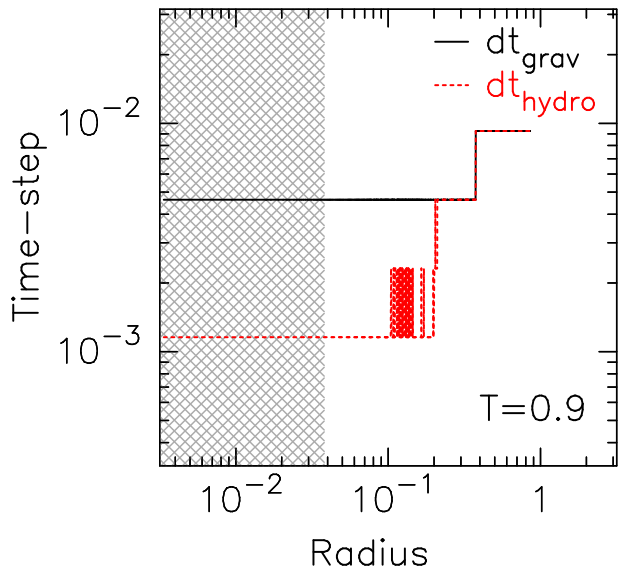}
\caption{
\label{fig:dts}
Radial profiles of $dt_{\rm hydro}$ and $dt_{\rm grav}$ for the spherical
collapse test with the individual time-step with FAST. The epoch is $T =
0.9$.  Solid and dotted curves indicate $dt_{\rm grav}$ and $dt_{\rm
hydro}$, respectively. The hatched region corresponds with the soften region
for the particle at the center by the gravitational softening.  
}
\end{center}
\end{figure}

\begin{figure}[htbp]
\begin{center}
\includegraphics[width=0.45 \textwidth]{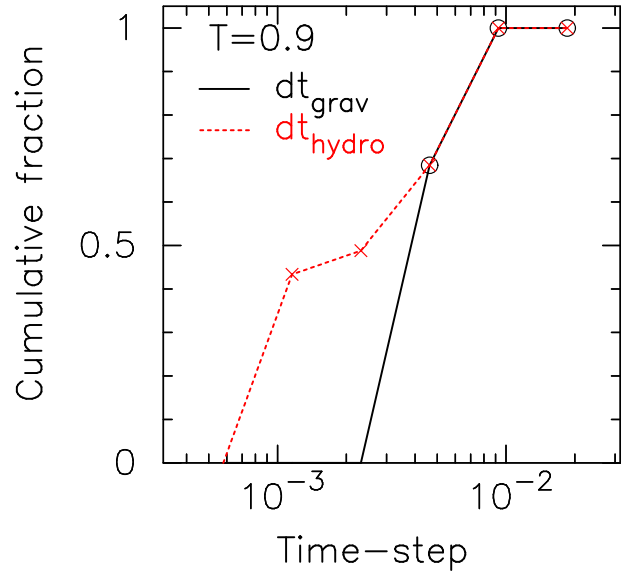}
\caption{
\label{fig:dts_fraction}
Cumulative fractions of particles as a function of $dt_{\rm hydro}$ and
$dt_{\rm grav}$ for the spherical collapse test with the individual
time-step with FAST. The epoch is $T = 0.9$.  Solid and dotted lines
indicate cumulative fractions of $dt_{\rm grav}$ and $dt_{\rm hydro}$,
respectively. 
}
\end{center}
\end{figure}

Figure \ref{fig:dts} shows the values of time-steps for gravity and
hydrodynamics for the run with the FAST method as a function of the distance
from the center at $T = 0.9$. Time-steps for gravity and hydrodynamics are
different in the post-shock region and the same in the ambient, pre-shocked
region. The transient region clearly matches with the shock front at the
radius of $\sim 0.2$ (see the top-left panel of figure
\ref{fig:3dcollapse}).  The values of $dt_{\rm hydro}$ and $dt_{\rm grav}$
in the post-shock region differ by a factor up to four.  Figure
\ref{fig:dts_fraction} shows cumulative fractions of $dt_{\rm hydro}$ and
$dt_{\rm grav}$ for the simulation with the individual time-step with FAST.
We can see that almost half of the particles have hydro time-steps smaller
than the minimum time-step for the gravity.

The errors of the total energy, the difference between the values of the
total energy at the initial ($T = 0$) and final ($T = 3$) states, for the
Ind and FAST methods are shown in table \ref{tab:3dcollapseErrors}.  $E_{\rm
s}$ and $E_{\rm f}$ represent the initial ($T=0$) total energy and the final
($T=3$) total energy, respectively.  These values are acceptably small.  In
our test runs, the absolute value of the energy error for the time
integration with FAST is smaller than that for the time integration without
FAST even though the time-step for gravity is larger.  The errors caused by
the gravitational and hydrodynamical integrations had the opposite signs and
partially canceled each other.

\begin{table}
\begin{center}
\caption{Total energy errors for the spherical collapse test.}\label{tab:3dcollapseErrors}
\begin{tabular}{lc}
\hline
\hline
Method & $|(E_{\rm s} - E_{\rm f})/E_{\rm s}|$ \\
\hline
Ind  & $3.0 \times 10^{-3}$ \\
FAST & $1.3 \times 10^{-3}$ \\
\hline
\end{tabular}\\
\end{center}
\end{table}

In table \ref{tab:3dcollapseTiming}, we show timing results of the collapse
test.  Our new method is faster than the original method, but not by a large
factor.  The calculation time of gravity is reduced to two thirds, and that
of tree construction is reduced to two fifths.  However, in this test the
hydrodynamics part dominates the total cost.

\begin{table}
\begin{center}
\caption{Timing results for the spherical collapse test.}\label{tab:3dcollapseTiming}
\begin{tabular}{lrrrr}
\hline
\hline
& \multicolumn{4}{c}{Time [sec]} \\
\cline{2-5} 
Method & Total & Gravity$^a$ & Hydro & Others \\
\hline
Ind  & $1754$ & $682~(52)$ & $960$ & $112$ \\
FAST & $1523$ & $468~(20)$ & $943$ & $112$ \\
\hline
\end{tabular}\\
\end{center}
$^a$ The tree structure construction times in the gravity part are shown as
parenthetical numbers.
\end{table}

Table \ref{tab:3dcollapseSteps} shows the number of steps and integrated
particles for the collapse test.  The ordinary individual time-step method
required $1979$ steps for the simulation in our implementation.  The FAST
method required $2022$ steps for the hydrodynamics part and $778$ steps for
the gravity part.  The reduction of the calculation time for tree
construction is directly proportional to the reduction of gravity steps.  In
our new method, the total number of integrated particles for the gravity
part becomes almost two thirds of that for the hydrodynamics part in this
test. 

\begin{table}
\begin{center}
\caption{Steps and number of integrated particles for the spherical collapse
test.}\label{tab:3dcollapseSteps}
\begin{tabular}{lrrrr}
\hline
\hline
& \multicolumn{2}{c}{Gravity} & \multicolumn{2}{c}{Hydro} \\
\cline{2-3} \cline{4-5}
Method & steps & $N_{\rm int,grav}$ & steps & $N_{\rm int,hydro}$ \\
\hline
Ind  & $1979$ & $2.4 \times 10^7$ & $1979$ & $2.4 \times 10^7$ \\
FAST & $778$  & $1.7 \times 10^7$ & $2022$ & $2.4 \times 10^7$ \\
\hline
\end{tabular}\\
\end{center}
\end{table}

\subsection{Test II: Three-dimensional explosion tests} \label{sec:explosion}

Now, we discuss the result of three-dimensional explosion test.  We designed
this test to mimic an explosion of a single SN in a self-gravitating gas
cloud.  When we take the mass of $10^{6}~\Mo$ and the radius of $100~{\rm
pc}$ as typical values of a giant molecular cloud \citep{Dame+1986}, its
potential energy is $\sim 10^{51}~{\rm ergs}$ (here we assume the cloud is in
a virial equilibrium state). This value is comparable to the energy released
by a single Type II SN.  Therefore, in order to investigate the behavior of
an exploding cloud induced by SN, we solved the evolution of a gas cloud
with a positive total energy comparable to the absolute value of the
original total energy.

We used the particle distribution of the three-dimensional collapse
test at $T = 3$ as the initial particle distribution of this explosion test.
We added the thermal energy in the central 32 particles with SPH manner.
Since the original total energy of the system is $\sim -0.6$, the new total
energy of the system was set to be $E = 0.5, 1, 2$, and $10$.  We set $T$ to
zero before the first step of the explosion calculation and follow the
evolution to $T = 5$.  In this test, we also plot the results of the global
time-step case.

\begin{figure*}[htbp]
\begin{center}
\includegraphics[width=0.9 \textwidth]{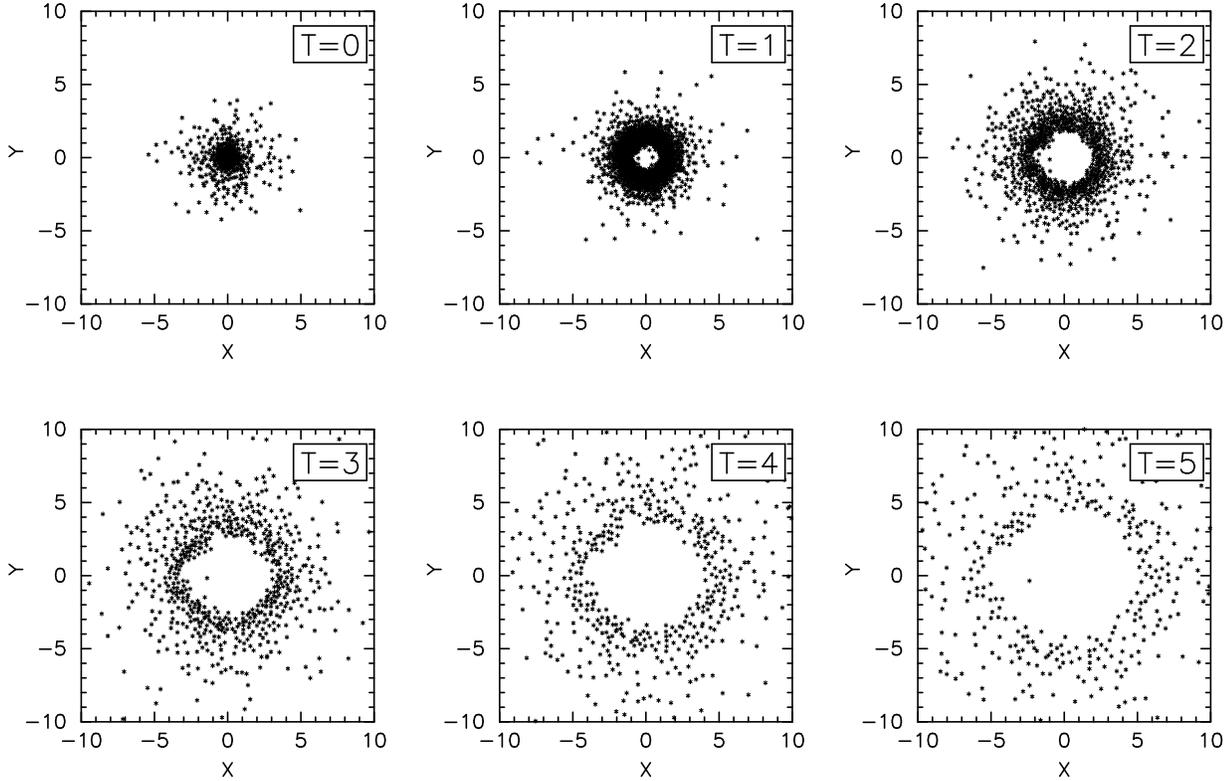}
\caption{
\label{fig:exp_slice}
Snapshots of the expanding cloud at six different epochs ($T = 0, 1,2,3,4$
and $5$). Projected particle distributions in a thin ($|z| < 0.1$) region
are shown. Dots indicate projected particle positions.  This is the case
that the explosion simulation with $E=2$.  The time-integration was done by
the individual time-steps with the FAST method.
}
\end{center}
\end{figure*}

Figure \ref{fig:exp_slice} shows snapshots of the expanding cloud, for the
case of $E = 2$, obtained with the FAST method at six different epochs ($T =
0, 1, 2, 3, 4$ and $5$). The particles in the region $|z| < 0.1$ are shown
in this figure.  The initially compact gas cloud expands driven by the high
pressure gas added in the center of the cloud, and forms a spherical
shell-like structure.  The shell moves outward, and at the final phase ($T =
5$), the radius of the shell becomes $\sim 5$.

\begin{figure}[htbp]
\begin{center}
\includegraphics[width=0.45 \textwidth]{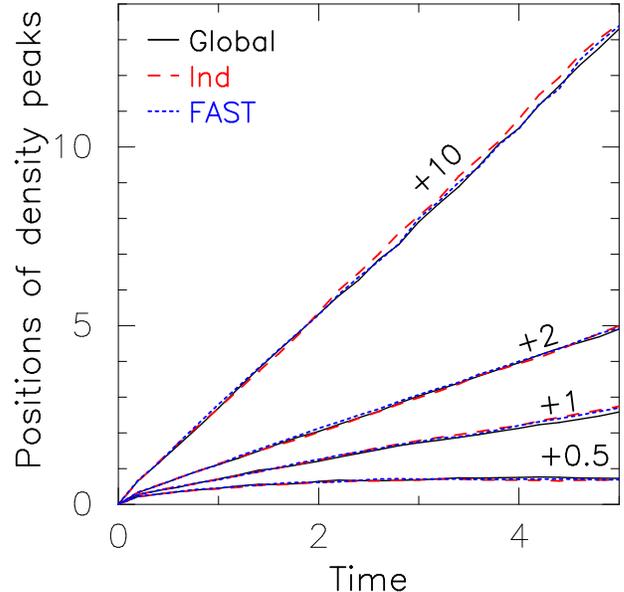}
\caption{
\label{fig:peakevolution}
Positions of density peaks as a function of time for various total energy
cases. Solid, dashed, and dotted lines indicate evolutions of density peaks
for cases of the global time-step (Global), ordinary individual time-steps
(Ind), and individual time-steps with FAST (FAST).  Numbers just above the
lines indicate the values of the total energy.
}
\end{center}
\end{figure}

Figure \ref{fig:peakevolution} shows time evolutions of density peaks for
simulations with several different values of the injection energy.  The
positions of peaks are derived by averaging the positions of the 10
particles with highest local density.  For the reference, in this figure, we
plotted the results obtained with the global-step method.  There are good
agreements between individual time-steps with/without FAST runs and the
global time-step runs.  This is because we adopted the time-step limiter for
hydrodynamics \citep{SaitohMakino2009}. Without this limiter, we would have
failed to obtain agreements between different methods.  The difference of
the the positions between individual time-steps with/without FAST runs are
summarized in table \ref{tab:diff}.  $R_{\rm Ind}$ and $R_{\rm FAST}$
represent radii of shells at $T = 5$ for individual time-step without/with
FAST and global time-step runs, respectively.  In this table, we also show
the difference between $R_{\rm FAST}$ and $R_{\rm Global}$, which is the
radius of the shell at $T = 5$ for the global time-step runs.  The
difference between FAST and Ind is comparable or smaller than the difference
for the result of global time-step.

\begin{table*}
\begin{center}
\caption{Differences between the peak positions with individual time-step
with FAST and others (individual time-step without FAST and global time-step).
}\label{tab:diff}
\begin{tabular}{lcccc}
\hline
\hline
& $E=0.05$ & $E=0.1$ & $E=0.2$ & $E=10$ \\
\hline
$| \frac{R_{\rm FAST}-R_{\rm Ind}}{R_{\rm Ind}}|$ & 
    $1.5~$\% & $1.6~$\% & $1.3~$\% & $0.4~$\% \\
$|\frac{R_{\rm FAST}-R_{\rm Global}}{R_{\rm Global}}|$ & 
    $6.4~$\% & $6.2~$\% & $1.9~$\% & $1.1~$\% \\
\hline
\end{tabular}\\
\end{center}
\end{table*}

Table \ref{tab:explosion:Timing} shows the timing results for the explosion
test.  We can see that the reduction in the cost of gravity calculation is
much larger than that in the collapse test, and the reduction in the cost of
tree construction is even larger. For $E = 10$, reduction in the tree
construction cost is a factor of eight.

\begin{table}
\begin{center}
\caption{Timing results for cloud explosion tests.}
\label{tab:explosion:Timing}
\begin{tabular}{lcrrrr}
\hline
\hline
& & \multicolumn{4}{c}{Time [sec]} \\
\cline{3-6} 
Method & $E$ & Total & Gravity$^a$ & Hydro & Others \\
\hline
Ind    & $0.5$ & $1264$ & $498~(138)$  & $550$  & $216$ \\
FAST   & $0.5$ & $972$  & $175~(16)$   & $568$  & $229$ \\
\hline
Ind    & $1$ & $1034$ & $396~(100)$ & $458$  & $180$ \\
FAST   & $1$ & $799$  & $135~(15)$  & $482$  & $182$ \\
\hline
Ind    & $2$ & $941$  & $358~(92)$  & $438$  & $145$ \\
FAST   & $2$ & $719$  & $112~(14)$  & $451$  & $156$ \\
\hline
Ind    & $10$ & $1018$ & $383~(106)$ & $466$  & $169$ \\
FAST   & $10$ & $758$  & $89~(13)$   & $484$  & $185$ \\
\hline
\end{tabular}\\
\end{center}
$^a$ Items are the same as table \ref{tab:3dcollapseSteps}.
\end{table}

The number of integrated particles and steps are summarized in table
\ref{tab:explosion:steps}.  By using the FAST method, we can greatly reduce
steps for the gravity part.  In these simulations, the use of the FAST
method resulted in the reduction of the number of gravity steps by a factor
of $7$ to $9$.  The numbers of integrated particles for the gravity part are
reduced to only $0.5-0.7$ times that of Ind runs.  As is shown above, the
speed up factors for the gravity part are $2.8-4.3$.  These results indicate
that the decrease of the number of steps is quite efficient for integrations
of self-gravitating fluid.  There is almost no change in the hydrodynamics
part.

\begin{table}
\begin{center}
\caption{Steps and number of integrated particles for cloud explosion tests.}
\label{tab:explosion:steps}
\begin{tabular}{lcrrrr}
\hline
\hline
& & \multicolumn{2}{c}{Gravity} & \multicolumn{2}{c}{Hydro} \\
\cline{3-4} \cline{5-6}
Method & $E$ & steps & $N_{\rm int,grav}$ & steps & $N_{\rm int,hydro}$ \\
\hline
Ind    & $0.5$ & $4675$ & $9.0 \times 10^6$  & $4675$ & $9.0 \times 10^6$  \\
FAST   & $0.5$ & $525$  & $5.9 \times 10^6$  & $4431$ & $8.8 \times 10^6$ \\
\hline
Ind    & $1$ & $3480$ & $7.8 \times 10^6$  & $3480$ & $7.8 \times 10^6$  \\
FAST   & $1$ & $479$  & $4.8 \times 10^6$  & $3535$ & $7.8 \times 10^6$ \\
\hline
Ind    & $2$ & $3515$ & $7.5 \times 10^6$ & $3515$ & $7.5 \times 10^6$  \\
FAST   & $2$ & $452$  & $4.1 \times 10^6$ & $3397$ & $7.4 \times 10^6$  \\
\hline
Ind    & $10$ & $3104$ & $7.5 \times 10^6$  & $3104$ & $7.5 \times 10^6$  \\
FAST   & $10$ & $427$  & $3.4 \times 10^6$  & $3412$ & $7.6 \times 10^6$ \\
\hline
\end{tabular}\\
\end{center}
\end{table}

\subsection{Test III: Merger simulations} \label{sec:merger}

In this section, we discuss the result of the application of the FAST method
to a realistic problem, namely simulations of galaxy-galaxy collisions.
Simulations we performed here were based on our recent galaxy-galaxy
collision simulations of \citet{Saitoh+2009}, in which we followed the
cooling of gas down to $10~{\rm K}$. We used the model of M1C.  Gravity,
hydrodynamics, radiative cooling, far-ultraviolet heating, star formation,
and type-II SNe were taken into account. The condition for the star
formation is that the gas is dense ($n_{\rm H} > 100~{\rm cm^{-3}}$) and
cold ($T < 100~{\rm K}$) with converging flows.  The regions which satisfy
these conditions form stars following the Schmidt-law with the local
star-formation efficiency of 0.033.  Further details of the modeling of star
formation were described in \citet{Saitoh+2008} and \citet{Saitoh+2009}.
Gravitational softening was set to be $20~{\rm pc}$ for all particles.  The
initial numbers of dark matter, (old) star, and gas particles were 6930000,
341896, and 148104, respectively.  We used 128 cores of Cray XT4 system at
Center for Computational Astrophysics of National Astronomical Observatory
of Japan.

\begin{figure}[htbp]
\begin{center}
\includegraphics[width=0.45 \textwidth]{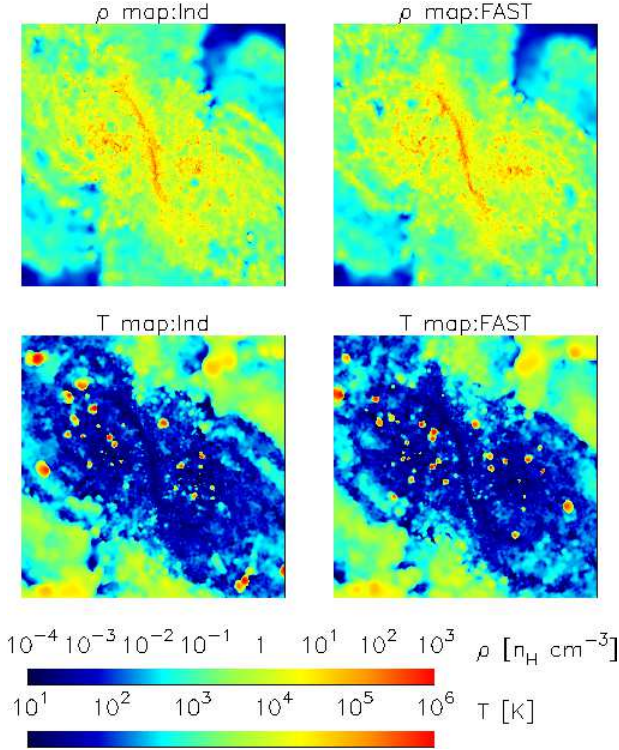}
\caption{
\label{fig:merger:snapshot}
Density and temperature maps for simulations by individual time-steps
without and with the FAST method.  Each panel shows $16~{\rm kpc} \times
16~{\rm kpc}$ in the orbital plane.  The left and right columns show the
results by individual time-steps without and with the FAST method,
respectively.  The epoch of these maps is $T = 420~{\rm Myr}$.
}
\end{center}
\end{figure}

Figure \ref{fig:merger:snapshot} shows density and temperature maps for
merger simulations at $T = 420~{\rm Myr}$ by individual time-steps without
and with the FAST method.  We can easily see that these two integration
methods show quite similar results in density and temperature structures.
The positions of ``Heat spots'' due to SNe are not perfectly identical
because of run-to-run fluctuations.  Other global properties of these
galaxies are also identical for both runs.

\begin{figure}[htbp]
\begin{center}
\includegraphics[width=0.45 \textwidth]{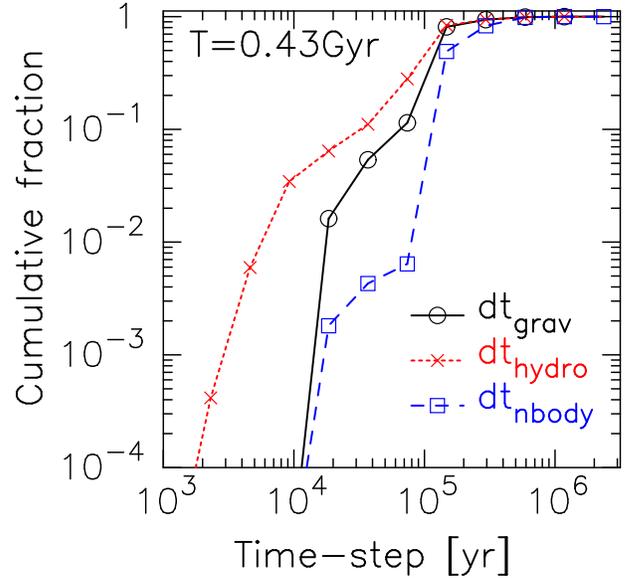}
\caption{
\label{fig:merger:dts_fraction}
Cumulative fractions of particles as a function of $dt_{\rm hydro}$ and
$dt_{\rm grav}$ for SPH particles and $dt_{\rm nbody}$ for collisionless
particles.  The time-steps are sampled from the merger simulation by the
individual time-step with FAST. The epoch is $T = 0.43~{\rm Gyr}$.  Solid,
dotted, and dashed histograms indicate cumulative fractions of $dt_{\rm
grav}$ and $dt_{\rm hydro}$ for SPH particles and $dt_{\rm nbody}$ for
collisionless particles.  
}
\end{center}
\end{figure}

Figure \ref{fig:merger:dts_fraction} shows cumulative fractions of particles
as a function of $dt_{\rm hydro}$ and $dt_{\rm grav}$ for SPH particles and
$dt_{\rm nbody}$ for collisionless particles.  The minimum time-step for the
hydrodynamics part is shorter than that for the gravity part for SPH
particles by a factor of eight.  In addition, collisionless particles have
longer time-steps than SPH particles.  Therefore the gravity part is skipped
in the lowest $3$ levels and the calculation is accelerated significantly.

Table \ref{tab:merger:Timing} shows timing results of merging simulations.
We sampled two typical epochs, {\it i.e.}, $350~{\rm Myr} \le T < 400~{\rm
Myr}$ and $400~{\rm Myr} \le T < 450~{\rm Myr}$.  The former epoch is a
quiescent star forming phase before the first encounter while the later
epoch is a significantly enhanced star forming phase during the first
encounter.  The total integration time for the simulation with the FAST
method decreases by almost a factor of two from that of the simulation
without the FAST method.  With the FAST method, the gravity part is $\sim 7$
times faster than that without the FAST method.  The reduction of the total
calculation time is similar for quiescent and starburst phases.

\begin{table*}
\begin{center}
\caption{Timing results for merger simulations.}
\label{tab:merger:Timing}
\begin{tabular}{lcrrrr}
\hline
\hline
& & \multicolumn{4}{c}{Time [sec]} \\
\cline{3-6} 
Method & Epoch & Total & Gravity$^a$ & Hydro & Others$^b$ \\
\hline
Ind  & $350~{\rm Myr} \to 400~{\rm Myr}$ & $14301$ & $5887~(2132)$  & $3545$  & $4869$ \\
FAST & $350~{\rm Myr} \to 400~{\rm Myr}$ & $7249$ & $919~(264)$  & $2923$  & $3407$ \\
\hline
Ind  & $400~{\rm Myr} \to 450~{\rm Myr}$ & $16454$ & $6703~(2441)$  & $4041$  & $5710$ \\
FAST & $400~{\rm Myr} \to 450~{\rm Myr}$ & $9646$ & $953~(279)$ & $4152$  & $4541$ \\
\hline
\end{tabular}\\
\end{center}
$^a$ Items are the same as table \ref{tab:3dcollapseSteps}.
$^b$ In this test runs, ``Others'' includes the calculation times of the
radiative cooling, star formation, and SNe routines.
\end{table*}

In table \ref{tab:merger:steps}, we show numbers of time-steps and
integrated particles for gravity and hydrodynamics parts. The simulation
with the FAST method required $\sim 7$ times smaller number of gravity steps
than that without the FAST method. Note that the number of integrated
particles for gravity is almost the same for the Ind and FAST method. The
large reduction in the calculation time is due to both the reduction of the
number of tree constructions and the removal of force calculations with
small number of particles, where the calculation becomes inefficient in
individual time-steps or parallel computers.

\begin{table*}
\begin{center}
\caption{Steps and number of integrated particles for merger simulations.}
\label{tab:merger:steps}
\begin{tabular}{lcrrrrr}
\hline
\hline
& & \multicolumn{2}{c}{Gravity} & \multicolumn{2}{c}{Hydro} & \\
\cline{3-4} \cline{5-6}
Method & Epoch & steps & $N_{\rm int,grav}$ & steps & $N_{\rm int,hydro}$ \\
\hline
Ind  & $350~{\rm Myr} \to 400~{\rm Myr}$ & $18697$ & $1.6 \times 10^9$  & $18697$ & $5.8 \times 10^7$ \\
FAST & $350~{\rm Myr} \to 400~{\rm Myr}$ & $2310$  & $1.6 \times 10^9$  & $16626$ & $6.4 \times 10^7$ \\
\hline
Ind  & $400~{\rm Myr} \to 450~{\rm Myr}$ & $21296$ & $1.8 \times 10^9$  & $21296$ & $7.3 \times 10^7$ \\
FAST & $400~{\rm Myr} \to 450~{\rm Myr}$ & $2425$  & $1.8 \times 10^9$  & $22959$ & $8.0 \times 10^7$ \\
\hline
\end{tabular}\\
\end{center}
\end{table*}

\section{Maximum Acceleration Factor by FAST} \label{sec:Discussion}
In this section, we estimate the maximum acceleration factor due to the
introduction of the FAST method using a simple calculation cost model.  By
comparing the calculation costs of the runs with/without FAST, we obtain the
acceleration factor due to the FAST method.

We here model the calculation cost of a simulation as follow:
\begin{eqnarray}
t_{\rm all} &=& N_{\rm s,g} (t_{\rm t,g} + \tilde{N}_{\rm u,g} t_{\rm e,g}) 
+ N_{\rm s,h} (t_{\rm t,h} + \tilde{N}_{\rm u,h} t_{\rm e,h}), \label{eq:tall} \\
&=& N_{\rm s,g} t_{\rm grav} + N_{\rm s,h} t_{\rm hydro}. \label{eq:tall_simple}
\end{eqnarray}
where $N_{\rm s,g}$ and $N_{\rm s,h}$ are the number of steps for gravity
and hydrodynamics, $t_{\rm t,g}$ and $t_{\rm t,h}$ are the calculation times
of tree construction for gravity and hydrodynamics, $\tilde{N}_{\rm u,g}$
and $\tilde{N}_{\rm u,h}$ are the mean number of updated particles for
gravity and hydrodynamics in each step, $t_{\rm e,g}$ and $t_{\rm e,h}$ are
the mean evaluation time of gravity and hydrodynamical interactions for a
single particle, respectively, and $t_{\rm grav}$ and $t_{\rm hydro}$ are
the mean calculation times in a single step for gravity and hydrodynamics,
respectively. In this model, we neglected the calculation cost of
miscellaneous operations, such as domain decomposition, time-integration,
evaluations of time-steps.  

In traditional individual time-steps, the number of gravity steps is the
same as that of hydrodynamical steps. Therefore the total calculation cost
with traditional individual time-steps is 
\begin{equation}
t_{\rm all,Ind} = N_{\rm s,h,Ind} (t_{\rm grav,Ind} + t_{\rm hydro,Ind}). \label{eq:tall_I}
\end{equation}
On the other hand, in FAST, the number of gravity steps is different from
that of hydrodynamical steps.  According to the argument in section
\ref{sec:Estimation}, the number of hydrodynamical steps is about ten times
larger than the gravity steps. It means $N_{\rm s,h,FAST} = 10 \times N_{\rm
s,g,FAST}$ in equation (\ref{eq:tall_simple}).  Thus, the total calculation
cost with FAST expresses as 
\begin{equation}
t_{\rm all,FAST} = N_{\rm s,h,FAST} 
(\frac{1}{10} t_{\rm grav,FAST} + t_{\rm hydro,FAST}). \label{eq:tall_F}
\end{equation}

The acceleration factor is defined as $\tau = {t_{\rm all,Ind}}/{t_{\rm all,FAST}}$.
If we assume that the number of hydrodynamical steps in the calculation without
FAST is the same as that with FAST and that calculation times between
with/without FAST are the same, the acceleration
factor is 
\begin{equation}
\tau = \frac{10 (t_{\rm grav} + t_{\rm hydro})}{t_{\rm grav} + 10 t_{\rm hydro}}. \label{eq:tau}
\end{equation}
In this equation, we removed suffixes ${\rm Ind}$ and ${\rm FAST}$ of
calculation costs.  Here we consider three typical cases that $t_{\rm grav}
\gg t_{\rm hydro}$, $t_{\rm grav} = t_{\rm hydro}$, and $t_{\rm grav} \ll
t_{\rm hydro}$.  The first case corresponds to usual $N$-body/SPH
simulations. In this case, the acceleration factor is $\tau = 10$, and is
quite large.  The second case corresponds to the case that (a) the
calculation time of the gravitational force reduces significantly by
adopting hardware/software accelerators, such as GRAPE or Phantom-GRAPE,
and/or (b) the calculation cost of hydrodynamics is rather expensive because
of the treatment of complex baryon physics, for instance star formation,
SNe, and chemical evolution.  The acceleration factor in this case is
$\tau \sim 2$.  Our simulations are close to the second case.
The final case is the ideal case that the calculation cost for gravity is
negligible.  In this case, the acceleration factor becomes unity and the gain
due to FAST is zero. We do not think this hypothetical situation can occur.

\section{Summary} \label{sec:Summary}

In this paper, we describe a fast integrated method, ``FAST'' for
self-gravitating fluid.  The FAST method assigns different time-steps for
gravitational and hydrodynamical interactions and integrates them
asynchronously.  The formulation of the FAST method is similar to the multi
time-step method \citep{Streett+1978} and also regarded as an extension of
``multi-step'' symplectic integrators, such as mixed variable symplectic
\citep{WisdomHolman1991}, multiple stepsize \citep{SkeelBiesiadechi1994},
and the BRIDGE \citep{Fujii+2007} methods.

The approach of the FAST method is qualitatively different from other
reduction techniques of the tree construction in gravity part. Thus the FAST
method eliminates unnecessary tree constructions and gravity calculations by
adopting longer time-steps for gravitational evolution than that for
hydrodynamics.

We found that the evolution of collapsing and exploding self-gravitating
fluid by the FAST method are identical to these by the usual unsplit method
which integrates gravity and hydrodynamics synchronously.

As a realistic test, we applied the FAST method to merger simulations
including self-gravity, hydrodynamics, radiative cooling, far-ultraviolet
heating, and SN \citep{Saitoh+2009}. In this test, we found that simulations
with and without the FAST method showed quite similar evolution.  The
calculation with FAST was nearly a factor of two faster.  This large gain
was due to the reduction in the gravity steps with small number of
particles.  The FAST method is very effective in accelerating simulations of
self-gravitating fluid.

\bigskip
We thank the anonymous referee for his/her insightful comments and
suggestions, which helped us to greatly improve our manuscript. 
We also thank Takashi Ito, Keiichi Wada, Michiko Fujii and Tomoaki Ishiyama for
useful discussion and Kohji Yoshikawa, who kindly provided us
with a custom version of the Phantom-GRAPE library.  A part of numerical
tests were carried out on Cray XT4 and GRAPE system at Center for
Computational Astrophysics of National Astronomical Observatory of Japan.
This project is supported by Grant-in-Aid for Scientific Research (17340059)
of JSPS, MEXT Japan the Special Coordination Fund for Promoting Science and
Technology, ``GRAPE-DR Project'', and Molecular-Based New Computational
Science Program of NINS.  TRS is financially supported by a Research
Fellowship from the Japan Society for the Promotion of Science for Young
Scientists.

\bigskip
\appendix 

\section{Symplectic integration method and its variants} \label{sec:SymplecticSchemes}

In this appendix, we explain the symplectic integration method briefly.
Then we explain sophisticated versions of symplectic integration methods,
{\it i.e.}, mixed variable symplectic method \citep{WisdomHolman1991},
multiple stepsize method \citep{SkeelBiesiadechi1994}, and the BRIDGE
\citep{Fujii+2007} methods.

Symplectic integration methods (e.g., \cite{DragtFinn1976, ForestRuth1990,
Yoshida1990, Yoshida1993})  are now widely used in the simulations of
gravitating $N$-body systems.  These methods preserve the symplectic form
of the canonical equation of motions when calculating the time variation of
the system.  This character leads to the very good conservation of system's
total energy on the course of numerical calculation.

When we express $H$ as the Hamiltonian of the system, and $p$ and $q$ as
six-dimension coordinates, canonical equations are 
\begin{eqnarray}
\frac{dq}{dt} &=& \frac{\partial H}{\partial p}, \\
\frac{dp}{dt} &=& -\frac{\partial H}{\partial q}.
\end{eqnarray}
We can summarize above two equations as
\begin{equation}
\frac{d f}{dt} = \{ f,H \}, \label{eq:poisson}
\end{equation}
where $f$ is $p$ or $q$, respectively, and $\{,\}$ is a Poisson bracket.  We
define an operator that 
\begin{equation}
\{ ,H \} f \equiv \{ f,H \}. \label{eq:poisson2}
\end{equation}
We can write a generalized canonical equation (\ref{eq:poisson}) as
\begin{equation}
\frac{d f}{dt} = \{ ,H \} f. \label{eq:poisson3}
\end{equation}
When we integrate equation (\ref{eq:poisson3}) from $t$ to $t + \Delta t$,
the formal solution of the equation (\ref{eq:poisson3}) is written as
\begin{equation}
f(t+\Delta t) = e^{\Delta t \{,H \}} f(t). \label{eq:integral} 
\end{equation}

Here, we consider the Hamiltonian of a self-gravitating system with $N$
particles.  In this case, the Hamiltonian is written as 
\begin{equation}
H = H_{\rm A} + H_{\rm B},
\end{equation}
where 
\begin{eqnarray}
H_{\rm A} &=& \sum_{i}^{N} \frac{p_i^2}{2 m_i},\\
H_{\rm B} &=& - \sum_{i<j}^{N} \frac{G m_i m_j}{q_{ij}}.
\end{eqnarray}
The formal solution is written as
\begin{equation}
f(t+\Delta t) = e^{\Delta t ( \{,H_{\rm A}\} + \{, H_{\rm B} \})} 
    f(t). \label{eq:integral2}
\end{equation}

Applying Barker-Champbell-Hausdorff formula \citep{Varadarajan1984book} to
equation (\ref{eq:integral2}), we obtain a first order integrator   
\begin{equation}
f(t+\Delta t) \approx e^{\Delta t \{,H_{\rm A}\} } 
    e^{\Delta t \{, H_{\rm B} \}} f(t), \label{eq:firstorder}
\end{equation}
and a second order integrator 
\begin{equation}
f(t+\Delta t) \approx e^{\frac{\Delta t}{2}\{,H_{\rm B}\} } 
    e^{\Delta t \{, H_{\rm A} \}} 
    e^{\frac{\Delta t}{2} \{,H_{\rm B}\} } f(t). \label{eq:secondorder}
\end{equation}
This second order integrator is well known as the {\it leap-frog}
integrator.

It is widely known that symplectic integration methods can achieve
high-accuracy once we split Hamiltonians into several components.   Mixed
variable symplectic (MVS) method \citep{WisdomHolman1991, Kinoshita+1991}
splits the Hamiltonian into an unperturbed part with an analytic solution
(i.e., Keplerian motion when we traced planetary motion) and a perturbation
part (i.e., mutual gravitational perturbation among planets).  When the
system is nearly integrable, the Hamiltonian for the unperturbed part
becomes much larger than that of the perturbed part, which enables the
method to accomplish a very high accuracy in integrating the equations of
motion, compared with conventional symplectic integrators.

The multiple stepsize (MSS) method \citep{SkeelBiesiadechi1994, Duncan+1998}
splits a potential into the sum of potentials of a short-range and a
long-range forces and gives different time-steps for different ranges of
forces.  The MSS method accomplishes the similar accuracy compared with the
usual symplectic method applied with small time-step.  The use of different
time-steps for different interactions was proposed for the integration of
molecular dynamics \citep{Streett+1978}. {\tt GADGET-2}
\citep{Springel2005}, which employs a TreePM method for gravitational force
calculation, adopts different time-steps for the long-range force derived
from a Particle-mesh method \citep{HockneyEastwood1981} and the short-range
force derived from a Tree method \citep{BarnesHut1986}.

BRIDGE \citep{Fujii+2007} was developed in order to solve galaxy-star
cluster systems self-consistently.  BRIDGE divides a Hamiltonian of a
galaxy-star cluster system into a star cluster and a galaxy parts, and
applies different time-steps and integrators.  In BRIDGE, the integration of
star cluster particles is performed by a forth-order integration method,
namely Hermit method \citep{MakinoAarseth1992}.  Force calculations among
star clusters are performed by direct method with GRAPE
\citep{Sugimoto+1990}.  The integration of galaxy particles is performed by
the leap-frog method with the Tree method for the force estimation.  Forces
between star cluster particles and galaxy particles are also calculated by
the Tree method with constant time-step.  Therefore this method can deal
with coevolution of collisionless and collisional systems self-consistently.


\end{document}